\documentclass[12pt]{article}
\usepackage{amsmath,amssymb,amsfonts,amsthm}
\title{On a general formalism of nonlinear charge coherent states,
their quantum statistics and nonclassical properties}

 \author{F. Eftekhari and M. K. Tavassoly
\\
\footnotesize{Atomic and Molecular Group, Faculty  of Physics,
Yazd University, Yazd, Iran}
\\ \footnotesize{e-mail: mktavassoly@yazduni.ac.ir  } }
\begin{document}

\date{\today}
\newcommand{\I}{\mathbb{I}}
\newcommand{\norm}[1]{\left\Vert#1\right\Vert}
\newcommand{\abs}[1]{\left\vert#1\right\vert}
\newcommand{\set}[1]{\left\{#1\right\}}
\newcommand{\R}{\mathbb R}
\newcommand{\C}{\mathbb C}
\newcommand{\eps}{\varepsilon}
\newcommand{\To}{\longrightarrow}
\newcommand{\BX}{\mathbf{B}(X)}
\newcommand{\HH}{\mathfrak{H}}
\newcommand{\D}{\mathcal{D}}
\newcommand{\N}{\mathcal{N}}
\newcommand{\la}{\lambda}
\newcommand{\af}{a^{ }_F}
\newcommand{\afd}{a^\dag_F}
\newcommand{\afy}{a^{ }_{F^{-1}}}
\newcommand{\afdy}{a^\dag_{F^{-1}}}
\newcommand{\fn}{\phi^{ }_n}
 \newcommand{\HD}{\hat{\mathcal{H}}}

 \maketitle
 \begin{abstract}
   In this paper, we will present a general  formalism for constructing the nonlinear charge
   coherent states which in  special case lead to the standard charge coherent states.
   The $su_Q(1, 1)$ algebra as  a nonlinear deformed algebra realization
   of the introduced states is established. In addition,
   the corresponding even and odd nonlinear charge coherent states have been also introduced. The formalism
   has the potentiality to be applied to systems either with known "nonlinearity function" $f(n)$ or
   solvable quantum system  with known "discrete non-degenerate spectrum" $e_n$.
   As some physical appearances, a few known physical systems in the two mentioned
   categories have been considered. Finally, since the construction of nonclassical states is a central topic of quantum optics,
   nonclassical features and quantum statistical properties of the introduced
   states have been investigated by evaluating  single- and two-mode squeezing, $su(1, 1)$-squeezing,
   Mandel parameter and antibunching effect (via $g$-correlation function) as well as some of their
   generalized forms we have introduced in the present paper.
 \end{abstract}



\section{Introduction}\label{sec-intro}

    Simultaneous with quantum mechanics birth,  Schr\"{o}dinger  while working on quantum harmonic oscillator,
    introduced coherent states \cite{schroedinger}.
    But it was further to prominent investigations of Glauber and Sudarshan
    \cite{gluber, sudarshan}
    on  genesis of laser as first accessible source of standard coherent stats
    that studies on these states  have been started, strictly.
    Many attempts in recent decades have been made along generalization of standard  coherent states.
    Nevertheless, all of the generalized coherent states can be displayed in three main categories.
    Based on algebraic ($q-$deformed \cite{biedenharn} and nonlinear coherent states \cite{matoes-filho, manko1997}),
    symmetric considerations (coherent states of $SU(1, 1)$ and $SU(2)$ group \cite{ barutan, perelomove})
    and finally the coherent
    states associated to potentials other than harmonic oscillator
    \cite{nieto-a, nieto-b, nieto-c}.
    However, in all of the above states  the quanta involved are uncharged.
    Pair coherent states were introduced in \cite{horn}.
    The usefulness of these states for the description of pion production in high-energy
    collisions is also clarified.
    But later they have been called as "charge coherent
    states" in \cite{bhaumik} and again as "pair coherent states" in \cite{gou-steinbach, AgarwalTara}.
    Any way, the notion of "charge coherent states" has proved it's successive usefulness
    in many areas of physics researches.
    As a few examples one may refer to elementary particles physics \cite{paricle1, paricle2},
    nuclear physics \cite{nuclear}, quantum physics \cite{qunt},
    quantum field theory \cite{quntf1, quntf2} and quantum optics \cite{quntopt1, quntopt2} fields.
    These states are generated through physical schemes in the quantum optics studies
    \cite{quntopt1, gen1, gen2, Botke}.
    The "standard charge coherent state"  are simultaneously eigenstates of two
    commuting operators, i.e., {\it charge operator}  given by
   \begin{equation}\label{charge-operator}
     Q=a^\dag_1a_1-a^\dag_2a_2,
   \end{equation}
     and {\it pair annihilation operator} $a_1a_2$ with  eigenvalues $q$ and
     $\xi$, respectively, i.e.,
  \begin{eqnarray}\label{ch-css}
   Q|\xi,q\rangle &=& q|\xi,q\rangle,\nonumber\\
   a_1a_2|\xi,q\rangle &=& \xi|\xi,q\rangle.
  \end{eqnarray}
  The explicit form of the eigenkets have been called as charge coherent states represent
  a well-known class of states within the general theory of
  coherent states and in the relevant literature \cite{horn2} read as
   \begin{eqnarray}\label{charge-cs}
    |\xi,q\rangle^{(+)} &=& \N_q^{(+)}(|\xi|^2)^{-1/2}\sum_{n=0}^\infty
    \frac {\xi^n}{\sqrt{n!(n+q)!}} |n+q, n \rangle ,\qquad q\geq 0,\nonumber \\
    |\xi,q\rangle^{(-)} &=& \N_q^{(-)} (|\xi|^2)^{-1/2} \sum_{n=0}^\infty \frac
    {\xi^n}{\sqrt{n!(n-q)!}} |n, n-q \rangle,\qquad q \leq 0,
   \end{eqnarray}
  where the kets $|m, n\rangle$ are the two-mode number states and
  $\N_q^{(\pm)}$ are the normalization factors may be determined as
   \begin{equation}\label{normcharge}
    \N_q^{(\pm)} = |\xi|^{\pm q}I_q(\mp 2 |\xi|)
   \end{equation}
   with $I_q(x)$ as the first order of modified Bessel function of
   the first kind.
   At this point it is worth to mention that from
   the group theoretical point of view the above states are nothing
   but a $SU(1, 1)$ coherent states. This is due to the fact that
   the operators $a_1a_2$ and it's conjugate are two of the three
   generators in the Schwinger-like definition of $SU(1, 1)$.
   In addition, the eigenvalues of the $Q$ charge operator is essentially
   the representation index of the principle-series representation of $SU(1, 1)$
   group where $C=(1-Q^2)/4$ represents the corresponding Casimir operator.
   The states in (\ref{charge-cs}) are simultaneously eigenstates of $k_-$ and $C$.


   Then, Liu constructed even and odd charge coherent states that are eigenstates
   of both operators  $Q$ and  $(a_1a_2)^2$  \cite{liu10}, i.e.,
  \begin{eqnarray}\label{e-o-ch-css}
    Q\;\;\;\;|\xi,q\rangle_{e(o)}^{(\pm)} &=& \;q|\xi,q\rangle_{e(o)}^{(\pm)},\nonumber\\
    (a_1a_2)^2|\xi,q\rangle_{e(o)} ^{(\pm)} &=& \xi^2|\xi,q\rangle_{e(o)}^{(\pm)}.
  \end{eqnarray}
   These latter states are  explicitly defined as
 \begin{equation}\label{charge-cs}
 |\xi,q\rangle^{(\pm)} = {\N'}_q^{(\pm)}(|\xi, q\rangle \pm |-\xi, q\rangle)
 \end{equation}
 where
 \begin{equation}
   {\N'}_q^{(\pm)} = \frac{1}{\sqrt 2}\left[\pm q  \pm (\N_q^{(\pm)})^2 |\xi|^{-q/2}J_q(\sqrt{2 |\xi|})\right]
 \end{equation}
  in which the closed form of $\N_q^{(\pm)}$ has been defined in (\ref{normcharge}) and $J_q(x)$
  is the Bessel function of the first kind.
  Liu {\it et al}   followed $k-$component coherent states idea in
  \cite{Sun} and constructed $k-$component charge coherent states
  which are  eigenstates of "charge" and the $k$th-powers of "pair boson annihilation" operators  \cite{liu13}.
  The well-known $\underline{q}-$deformed coherent  states \cite{biedenharn} are defined as the
  right eigenstates of $b=f_{\underline{q}}(n)a$, with
  \begin{equation}\label{f-qdeform}
  f_{\underline{q}}(n)=\sqrt{\frac{\underline{q}^{n+1}-\underline{q}^{-n-1}}{(n+1)
  (\underline{q}-\underline{q}^{-1})}},
  \end{equation}
  provided that  $0<\underline{q}\leq 1$. It is worth mentioning that the specific notation
  $\underline{q}$ in the description of $f_{\underline{q}}(n)$ is choosed here  to
   distinguish it from the  eigenvalues $q$ of $Q$ operator.
  The nonlinearity of $q$-deformed oscillators were realized in the dependence of frequency
  on the intensity ($n=a^\dag a$) of light.
  Using this deformation Chatourvedi and Srinivassan constructed
  $\underline{q}-$deformed charge coherent states that are
  eigenstates of both charge  operator in (\ref{charge-operator}) and pair $\underline{q}-$boson
  annihilation operators \cite{chaturvedi},
  $b_1 b_2$ with $b_i=f_{\underline{q}}(n_i)a_i$ for $i=1, 2$.
    The even and odd $\underline{q}-$deformed charge  coherent states are eigenstates of the square of
    the pair $\underline{q}-$deformed annihilation operators $(b_1b_2)^2$
    and the charge operator \cite{liu-quesene}.
    Liu by using the concept of $k-$component
    $\underline{q}-$deformed coherent states \cite{kuang} introduced the $k-$component
    $\underline{q}-$deformed charge coherent
    states \cite{liu-Q-song} that are eigenstates of charge operator in
    (\ref{charge-operator})  and $k$th-power of $\underline{q}-$deformed boson operator $(b_1 b_2)^k$.


   On the other side,  nonlinear coherent states were first introduced by de Matoes Filho and Vogel
   \cite{matoes-filho} and Man'ko {\it et al} \cite{manko1997}.
   These states are defined as right eigenstates of the $f-$deformed annihilation operators $A=af(n)$,
   where $f(n)$ is the nonlinearity function.
   As established by  Vogel  {\it et al} \cite{matoes-filho, vogel} a special set of these quantum
   states may be generated as the stationary states
   of the center-of-mass motion of a trapped ion far from the
   Lamb-Dicke regime.
   There has been shown that there are so many generalized coherent states that can be put in this important category
   with some special nonlinearity functions \cite{Roknizadeh2004, tavassoly-j.m.p}.
   Indeed, these states provide a powerful method which can unify the mathematical
   structure of a lot of generalized coherent states introduced  in the literature.
   So, a reasonable and natural extension of the standard and
   $\underline{q}$-deformed charge states is provided by the notion of
   "{\it $f$-deformed charge coherent states}".
   The main purpose of the present work is to introduce a general formalism for the construction of
   a wide classes of charge coherent states based on the nonlinear coherent states method.
   It is worth to mention that upon the relation between nonlinearity function of nonlinear
   coherent states and the Hamiltonian of the quantum systems proposed by one of us
   \cite{Roknizadeh2004, honarasa} the formalism can be extended to solvable quantum systems
   with discrete non-degenerate spectra. We hope that
   the introduced nonlinear charge coherent states in the present paper which stream to nonlinear physical systems
   may find their useful applications
   in various fields of researches, as the standard charge coherent states \cite{paricle1}-\cite{Botke}.

   The paper organized as follows. In section 2 we outline the general structure of nonlinear charge coherent states.
   The nonlinear algebraic realization of the $su_Q(1 ,1)$ generators for the introduced states
   is established in section 3.
   Then, by symmetric and antisymmetric superposition of nonlinear charge coherent states,
   even and odd nonlinear charge coherent states have been constructed in section 4. Next,
   in section 5  we study the nonclassical properties and quantum statistics of  the introduced states via investigating
   single- and two-mode squeezing in addition to $su(1, 1)$-squeezing,
   Mandel parameter,  and  two-mode second order correlation function together with all of their
    generalized forms will be introduced
   in the present work.
   Finally, in section 6 we apply
   our presented formalism to a few physical situations either solvable quantum systems with  known
   "discrete spectrum" $e_n$ or
   nonlinear oscillator systems with known "nonlinearity function" $f(n)$.
   For instance, "Sudarshan harmonious states" and  "$SU(1, 1)$ coherent states" as the first type and
    "hydrogen-like spectrum",
   "P\"{o}schl-Teller" and "infinite well potentials" as the second type will be considered and discussed.

 \section{Introducing the nonlinear  charge coherent states}\label{sec-nl}

   In this section using the nonlinear coherent states method \cite{manko1997, Roknizadeh2004}
   we introduce the nonlinear charge coherent states in a general structure.
   Consider $f-$deformed ladder operators
  \begin{equation}\label{f-ladder-operator}
    A_i=a_i f(n_i),\qquad\qquad {A^\dag}_i=f^\dag(n_i){a^\dag}_i,
  \end{equation}
   where $a_i$, $a_{i}^{\dag}$ and $n_i=a_i^\dag a_i,\quad i=1, 2$ are  respectively boson annihilation,
   creation  and number operators of $i$th mode, and  $f(n)$ is a nonlinear function characterizes
   the physical  systems. The pair $f-$deformed boson annihilation operator $A_1 A_2$ and charge operator
    commute with together, i.e.,
   $ \left[Q , A_1 A_2\right]=0,$
   where charge operator keeps its previous definition in (\ref{charge-operator}).
   Thus, the latter two operators  should satisfy eigenvalue equations as follow
  \begin{eqnarray}\label{nonlinear-eigenvalue}
    Q|\xi,q,f\rangle &=& q|\xi,q,f\rangle,\nonumber\\
    A_1A_2|\xi,q,f\rangle &=& \xi|\xi,q,f\rangle,
  \end{eqnarray}
    where $\xi \in \mathcal{C}$ and $q$ is an integer has been  called charge number. These
     eigenstates have generally the following expansion
  \begin{equation}\label{fock-form}
    |\xi,q,f\rangle=\sum_{n,m=0}^\infty c_{n,m}|n,m\rangle.
  \end{equation}
    Substituting  (\ref{fock-form}) in (\ref{nonlinear-eigenvalue}) and calculating expansion coefficients
    one straightforwardly obtains two distinct states as follows
  \begin{equation}\label{positive-non-ch-css}
    |\xi,q,f\rangle^{(+)} = {\mathcal{N}^{(+)}(|\xi|^2)}^{-1/2} \sum_{n=0}^{\infty} \xi^n
    \frac{1}{\sqrt{n!\left[n+q\right]!}\left[f(n)\right]! \left[f(n+q)\right]!}|n+q,n\rangle,\
    \quad q \geq 0,
  \end{equation}

  \begin{equation}\label{negative-non-ch-css}
    |\xi,q,f\rangle^{(-)}= {\mathcal{N}^{(-)}(|\xi|^2)}^{-1/2} \sum_{n=0}^{\infty}\xi^n
       \frac{\xi^n}{\sqrt{n!\left[n-q\right]!}\left[f(n)\right]! \left[f(n-q)\right]!}|n,n-q\rangle,\
    \quad q \leq 0.
  \end{equation}
     The latter states can be put in the following single expression
  \begin{eqnarray}\label{bouth-non-ch-css}
   |\xi,q,f\rangle&=&{\mathcal{N}(|\xi|^2)}^{-1/2}\sum_{n=0}^{\infty} \frac{\xi^n}
   {\sqrt{n!\left[n+|q|\right]!}\left[f(n)\right]! \left[f(n+|q|)\right]!}  \nonumber \\
    &\times&
    \biggl|n+\frac{q+|q|}{2},n-\frac{q-|q|}{2}\biggr\rangle,
  \end{eqnarray}
   with normalization constant given by
  \begin{equation}\label{nor-fac}
    \mathcal{N}(|\xi|^2)=\sum_{n=0}^{\infty}
    \frac{{|\xi|}^{2n}}{n!\left[n+|q|\right]!{\{\left[f(n)\right]!
    \left[f(n+|q|)\right]!\}}^2}.
  \end{equation}
    Note that in obtaining  (\ref{positive-non-ch-css})-(\ref{nor-fac}) using
    the conventional definitions
  \begin{equation}\label{cond-fn-fact}
       \left[f(n)\right]!\doteq f(n)f(n-1)...f(1),\qquad \qquad\left[f(0)\right]! \doteq 1,
  \end{equation}
   we have defined
  \begin{eqnarray}\label{cond-state}
    \left[n+|q|\right]!\doteq (n+|q|)(n-1+|q|)...(1+|q|),\quad &\left[|q|\right]! \doteq 1,&\nonumber\\
    \left[f(n+|q|)\right]!\doteq f(n+|q|)f(n-1+|q|)...f(1+|q|),\quad &\left[f(|q|)\right]!\doteq 1.&
  \end{eqnarray}
   To endorse our formalism we can substitute $f(n)=1$  in above relations and check that  standard charge coherent
   state in (\ref{charge-cs}) will be obtained.
   It is worth noticing that by substituting the $\underline{q}-$deformation function  (\ref{f-qdeform})
   in the above results one can not expect the exact result of $\underline{q}-$deformed charge coherent
   states in \cite{liu-quesene}. This is due to the fact that
   we began our presented formalism of nonlinear charge coherent states  with the deformed annihilation
   operator $A=af(n)$, while in  \cite{liu-quesene} the authors used the definition $A=f_{\underline{q}}(n)a$.
    Moreover, due to the relation  $f(n)a=af(n-1)$ replacing $n$ with $n-1$, the consistency
    of the formalism will be revealed.
    We would like to mention that the $q$-deformation
    nonlinearity in Eq. (8) has no zeroes at positive integers, and this feature is not
    changed by replacing $n$ with $n-1$. So in both cases the Ston-Von Numman theorem
    can be extended to the case of the related operators $A$ and $A^{\dag}$. Also the ladder operators are irreducible over the Fock space.

 \section{Deformed algebraic realization of the introduced states}

    Liu  considered the standard (undeformed) $su(1, 1)$ Lie algebra generators in terms
    of the two-mode boson operators as \cite{liu10}
  \begin{equation}\label{su11-qenerator}
    k_-=a_1a_2,\qquad k_+=a_{1}^{\dag} a_{2}^{\dag}, \qquad k_0=\frac 1 2 (a_{1}^{\dag} a_1+a_{2}^{\dag}
    a_2+1).
 \end{equation}
   The relavant coherent states are well-known charge coherent
   states or Barut-Girardello coherent states \cite{barutan, horn, bhaumik}.
   Keeping in mind the Liu work and following the path of Hu and Chen  in \cite{hu-chen},
   we choose the associated generators for  our  nonlinear
   charge coherent states using the $f-$deformed ladder operators in (\ref{f-ladder-operator}) as follows:
 \begin{equation}\label{nld-generator}
   K_-=A_1A_2,\quad K_+={A_1}^\dag{A_2}^\dag,\quad\ \left[K_-, K_+\right]=2 K_0,
 \end{equation}
   where $K_0$ can be calculated as
 \begin{equation}\label{k0}
   K_0=\frac{1}{2}[(n_1+1){f^2(n_1+1)}(n_2+1){f^2(n_2+1)}-n_1{f^2(n_1)}n_2{f^2(n_2)}].
 \end{equation}
    Obviously one has  $K_-=(K_+)^\dag$, $(K_-)^\dag=K_+$ and $K_0=K_0^\dag$. Now it can be easily checked that
     the following commutation  relations hold:
 \begin{equation}\label{commut-k0-k-}
    \left[K_0,K_-\right] = -K_-g(n_1,n_2),\qquad
    \left[K_0,K_+\right] = g(n_1,n_2)K_+,
 \end{equation}
      where we have defined:
 \begin{eqnarray}\label{g-n1-n2}
    g(n_1,n_2) & = & \frac 1 2 [ (n_1+1){f^2(n_1+1)}(n_2+1){f^2(n_2+1)}-n_1{f^2(n_1)} \nonumber\\
    &\times&n_2{f^2(n_2)} + (n_1-1){f^2(n_1-1)}(n_2-1){f^2(n_2-1)}].
 \end{eqnarray}
   Thus, we established the nonlinear deformed algebra denoted by us as $su_Q(1,1)$
    for the nonlinear charge coherent states in
   (\ref{bouth-non-ch-css}).

 \section{Even and odd nonlinear charge coherent states}

    We can construct the even and odd nonlinear charge coherent state via symmetric and antisymmetric
    superposition of nonlinear charge coherent states introduced in (\ref{positive-non-ch-css})
     and (\ref{negative-non-ch-css}) such as:
   \begin{eqnarray}\label{e-o-ch-css}
    |\xi,q,f\rangle_{e(o)} ^{(\pm)}&=&\frac 1 2 {\mathcal{N}^{(\pm)}(|\xi|^2)}^{1/2} \;\;
     {{\mathcal{N}^{(\pm)}_{e(o)}(|\xi|^2)}^{-1/2}} \nonumber\\  &\times& \left(|\xi,q,f\rangle ^{(\pm)} \pm |-\xi,q,f\rangle
     ^{(\pm)}\right),\qquad q \geq
    0.
   \end{eqnarray}
   The explicit form of the above states for the case $q\geq 0$ then straightforwardly may be obtained separately  as
   \begin{eqnarray}\label{e-non-gq}
    |\xi,q,f\rangle_e ^{(+)}&=&{\mathcal{N} ^{(+)}_e(|\xi|^2)}^{-1/2} \sum_{n=0}^\infty \xi^{2n}
    \nonumber\\  &\times&
    \frac{1}{\sqrt{(2n)!\left[2n+q\right]!}\left[f(2n)\right]!\left[f(2n+q)\right]!}
    |2n+q,2n\rangle,
  \end{eqnarray}
  \begin{eqnarray}\label{o-non-gq}
    |\xi,q,f\rangle_o ^{(+)} &=& {\mathcal{N} ^{(+)}_o(|\xi|^2)}^{-1/2}\sum_{n=0}^\infty
    \frac{\xi^{2n+1}}{\sqrt{(2n+1)!\left[2n+1+q\right]!}} \nonumber\\
   &\times&
    \frac{1}{\left[f(2n+1)\right]!\left[f(2n+1+q)\right]!}|2n+1+q,2n+1\rangle.
 \end{eqnarray}
    The two normalization constants  can be calculated as follows:
  \begin{equation}\label{norm-e-o}
    \mathcal{N}^{(+)}_{e(o)}(|\xi|^2)=
    \frac{1}{2}{\mathcal{N} ^{(+)}(|\xi|^2)}\pm\frac{1}{2}\sum_{n=0}^\infty\frac{(-|\xi|)^{n}}{n!\left[n
    +q\right]!{\{\left[f(n)\right]!\left[f(n+q)\right]!\}^2}},
  \end{equation}
    where $\mathcal{N}^{(+)}(|\xi|^2)$ may be obtained from (\ref{nor-fac}) when $q\geq 0$.
    By replacing $2n$ with $2n-q$ in relations
     (\ref{e-non-gq}) and (\ref{o-non-gq}) one can obtain the associated states for $q\leq 0$.  It is easy
     to check that these states may also be obtained as the eigenstates of squared of  pair $f-$deformed
      annihilation operators with eigenvalues ${\xi}^2$, i.e.,
   \begin{equation}\label{square-eigenvalue}
    (A_1A_2)^2|\xi,q,f\rangle_{e(o)}={\xi}^2|\xi,q,f\rangle_{e(o)}.
   \end{equation}
    Also, the following relations hold
   \begin{equation}\label{ortognal-charge}
     _{e(o)}\langle\xi,q,f|\xi,q',f'\rangle_{e(o)} = {\mathcal{N}_{e(o)}}({|\xi|}^2)^{-1/2}
     {\mathcal{N}_{e(o)}}({|\xi'|}^2)^{-1/2} {\mathcal{N}_{e(o)}}(|{\xi}^*\xi'|)\delta_{q,q'},
   \end{equation}

  \begin{equation}\label{ortognal-e-o}
   \qquad_e \langle\xi,q,f|{\xi}',q,f\rangle_o=0.
  \end{equation}
   The Dirac delta function in (\ref{ortognal-charge}) explains that the even  nonlinear charge coherent
   states are orthogonal to each other relative to charge number. Similar situation holds for odd
   nonlinear charge coherent states, too. In addition, the  relation (\ref{ortognal-e-o}) indicates that
   even and odd nonlinear charge coherent state with identical charge number and clearly
   the same nonlinearity function  are orthogonal.

 \section{Nonclassicality of the introduced states}\label{sec-n5-properties}

    Motivations to introduce generalized coherent states theoretically and to produce them in the laboratory
    are mainly due to their nonclassical properties which their usefulness in sensitive
    measurements is a well-known subject.
    In this section we would like to illustrate what nonclassical properties will be considered by us to
    investigate the nonclassicality features  of the introduced states.
    To achieve this purpose we will check various properties such as  single- and two mode-squeezing,
    $su(1, 1)$-squeezing, Mandel parameter and second order correlation function
    (also their new generalized forms which we will define).
    It is worth mentioning that only one of the latter
    properties is sufficient for a states to be considered as nonclassical states.
\subsection{Squeezing effects}\label{}
   Different squeezing parameters will be outlined in this subsection.

 \subsubsection{Single-mode squeezing and it's generalization}\label{subs-1-1-5}
    Usually quadrature operators of filed in two-mode are defined as follows
  \begin{equation}\label{y1-y2}
    y_1=\frac {{a_1}^\dag+a_1}{2},
    \qquad\qquad y_2=\frac {i({a_1}^\dag-a_1)}{2},
  \end{equation}
  \begin{equation}\label{z1-z2}
    z_1=\frac {{a_2}^\dag+a_2}{2},
     \qquad\qquad z_2=\frac {i({a_2}^\dag-a_2)}{2},
  \end{equation}
    where $y_i$,  $i=1, 2$, indicates to the quadratures of the first mode and $z_i$ to the second mode.
    They satisfy the canonical  commutation relations:
  \begin{equation}\label{comut-y1-z1}
     \left[y_1,y_2\right]=\frac{i}{2},
     \qquad\quad\left[z_1,z_2\right]=\frac{i}{2}.
  \end{equation}
   So, the uncertainty relation for above  conjugate operators read respectively as
  \begin{eqnarray}\label{unc-y1,z1}
     \langle(\Delta y_1)^2\rangle\langle(\Delta y_2)^2\rangle\geq\frac{1}{16},
     \quad\quad \langle(\Delta z_1)^2\rangle\langle(\Delta z_2)^2\rangle\geq\frac{1}{16},
  \end{eqnarray}
     where in these relations and all other cases which will be followed the uncertainties in any partner of a set
      of conjugate operator defined by $\langle (\Delta x)^2\rangle\doteq\langle x^2\rangle - \langle x\rangle^2$.
     Therefore,  single-mode squeezing occurs in the first mode  for any quantum state if
  \begin{equation}\label{con-squ-y1}
     \langle(\Delta y_i)^2\rangle<\frac{1}{4},\qquad i=1\; \mbox{or}\; 2,
  \end{equation}
     and similarly for the second mode.
     The corresponding fluctuations with respect to the states introduced in (\ref{bouth-non-ch-css}) are give by
  \begin{eqnarray}\label{squ-y1}
     \langle(\Delta y_1)^2\rangle &=& \langle(\Delta y_2)^2\rangle = \frac 1 4  \Biggl\{2 \mathcal{N}(|\xi|^2)^{-1}
      \nonumber\\  &\times& \sum_{n=0}^\infty
     \frac{{|\xi|}^{2n}}{(n-1)!\left[n+|q|\right]!
     \{\left[f(n)\right]!\left[f(n+|q|)\right]!\}^2}+ 1 \Biggr\},
  \end{eqnarray}
     and
  \begin{eqnarray}\label{squ-z1}
     \langle(\Delta z_1)^2\rangle &=& \langle(\Delta z_2)^2\rangle=  \frac 1 4  \Biggl\{2 \mathcal{N}(|\xi|^2)^{-1}
      \nonumber\\  &\times& \sum_{n=0}^\infty
     \frac{{|\xi|}^{2n}}{n!\left[n-1+|q|\right]!
     \{\left[f(n)\right]!\left[f(n+|q|)\right]!\}^2}+1 \Biggl\}.
  \end{eqnarray}
     Equations (\ref{squ-y1}) and (\ref{squ-z1}) show that the inequalities
     $\langle(\Delta y_i)^2\rangle\geq \frac{1}{4}$
      and  $\langle(\Delta z_i)^2\rangle\geq \frac{1}{4}$ always hold for $i=1\; \mbox{or}\; 2$.
     These results indicate that single-mode squeezing can not occur for any arbitrary function $f(n)$.

     Changing the boson  operators in relations (\ref{y1-y2}) and (\ref{z1-z2}) by $f-$deformed ladder operator,
     we define the "{\it generalized quadrature operators}" respectively for first and second mode  as follows
   \begin{equation}\label{G-y1-y2}
     Y_1=\frac {{A_1}^\dag+A_1}{2},
     \qquad\quad Y_2=\frac {i({A_1}^\dag-A_1)}{2},
  \end{equation}
  \begin{equation}\label{G-z1-z2}
     Z_1=\frac {{A_2}^\dag+A_2}{2},
     \qquad\quad Z_2=\frac {i({A_2}^\dag-A_2)}{2}.
   \end{equation}
     These two pairs of quadratures satisfy the noncanonical commutation relations:
   \begin{equation}\label{noncanon-commut-yb}
     \left[Y_1,Y_2\right]=\frac{i}{2}[(n_1+1){f^2(n_1+1)}-n_1{f^2(n_1)}],
  \end{equation}
  \begin{equation}\label{noncanon-commut-zb}
     \left[Z_1,Z_2\right]=\frac{i}{2}[(n_2+1){f^2(n_2+1)}-n_2{f^2(n_2)}].
  \end{equation}
     So, the uncertainty relations are respectively given by
   \begin{equation}\label{uncer-Yb}
      \langle(\Delta Y_1)^2\rangle\langle(\Delta Y_2)^2\rangle \geq \frac{1}{16}
      {\left|\langle(n_1+1){f^2(n_1+1)}-n_1{f^2(n_1)}\rangle\right|}^2,
  \end{equation}
  \begin{equation}\label{uncer-zb}
     \langle(\Delta Z_1)^2\rangle\langle(\Delta Z_2)^2\rangle\geq\frac{1}{16}
     {\left|\langle(n_2+1){f^2(n_2+1)}-n_2{f^2(n_2)}\rangle\right|}^2.
  \end{equation}
     A state is said to be single-mode squeezed in $Y_i$  if
      \begin{equation}\label{uncer-Yb}
      \langle(\Delta Y_i)^2\rangle  <  \frac{1}{4}{\left|\langle(n_1+1){f^2(n_1+1)}-n_1{f^2(n_1)}\rangle\right|}^2,
  \end{equation}
   and in $Z_i$ if
  \begin{eqnarray}\label{cond-squ-zb}
       \langle(\Delta Z_1)^2\rangle  <  \frac{1}{4}{\left|\langle(n_2+1){f^2(n_2+1)}-n_2{f^2(n_2)}\rangle\right|}^2.
  \end{eqnarray}
     where in (\ref{uncer-Yb}) and (\ref{cond-squ-zb}) $i=1\;\mbox{or}\;2$.
     It can be easily checked that
     single-mode squeezing can not hold for arbitrary $f(n)$
     function   similar to (\ref{squ-y1}) and (\ref{squ-z1}).

 \subsubsection{Two-mode squeezing and it's generalization}\label{sub-2-1-5}
     Following the definitions in \cite{liu10, loudon}  with the help of (\ref{y1-y2}) and (\ref{z1-z2}) we now
      introduce the two-mode hermitian quadrature operators:
  \begin{equation}\label{w1-w2}
     w_1=\frac{y_1+z_1}{\sqrt{2}},\qquad
     w_2=\frac{y_2+z_2}{\sqrt{2}},
  \end{equation}
     that satisfy the canonical  commutation relation
    \begin{equation}\label{commut-w1-w2}
      \left[w_1,w_2\right]=\frac{i}{2}.
   \end{equation}
      The uncertainty relation is simply obtained as
    \begin{equation}\label{uncer-w1-w2}
      \langle(\Delta w_1)^2\rangle\langle(\Delta w_2)^2\rangle\geq\frac{1}{16}.
   \end{equation}
      The two-mode squeezing for nonlinear charge coherent states may be observed if
   \begin{equation}\label{cond-squ-w1-w2}
      \langle (\Delta w_i)^2\rangle  < \frac{1}{4},\qquad\qquad i=1 \;\mbox{or} \; 2.
   \end{equation}
      The fluctuations in $w_1$ and $w_2$ may be calculated straightforwardly, where again
      it is observed that  there are not two-mode squeezing with respect to
      introduced states for any arbitrary $f(n)$ function.

     Generalizing the proposal, by replacing $a$ and $a^\dag$ respectively with $A$ and $A^\dag$ in relations
     (\ref{w1-w2})  we can define new "{\it generalized two-mode quadrature operators}"  as
    \begin{equation}\label{w1b-w2b}
      W_1 =\frac{Y_1+Z_1}{\sqrt{2}},\qquad
      W_2 = \frac{Y_2+Z_2}{\sqrt{2}},
    \end{equation}
      where $Y_i$ and $Z_i$ ($i=1, 2$) defined in (\ref{G-y1-y2}) and (\ref{G-z1-z2}).
      These  quadratures  satisfy the following commutation relation
   \begin{eqnarray}\label{commut-w1b-w2b}
    \left[W_1,W_2\right]&=& \frac{i}{4} [(n_1+1){f^2(n_1+1)}-n_1{f^2(n_1)}\nonumber\\
    &+&(n_2+1){f^2(n_2+1)}-n_2{f^2(n_2)}].
   \end{eqnarray}
      Uncertainty relation corresponding to these conjugate quadratures read as
   \begin{eqnarray}\label{uncer-w1b-w2b}
     \langle(\Delta W_1)^2\rangle\langle(\Delta W_2)^2\rangle & \geq & \frac{1}{4}
     \left|\langle\left[W_1,W_2\right]\rangle\right|^2.
   \end{eqnarray}
  So, generalized two-mode squeezing for nonlinear charge coherent states may be observed if
  \begin{equation}\label{cond-squ-w1b-w2b}
   S_{W_i}=\langle (\Delta W_i)^2 \rangle -  \frac{1}{2}|\langle\left[W_1,W_2\right]\rangle| < \; 0,
  \end{equation}
       where $i=1\;\mbox{or}\; 2$.

\subsubsection{$su(1, 1)-$squeezing and it's generalization} \label{sub-3-1-5}
     Following the Wodkiewicz and Eberly proposal in \cite{wodkiciwicz} we now consider two hermitian
     quadratures that constructed from usual $su(1, 1)$ generators $k_+$, $k_-$
     were defined in (\ref{su11-qenerator}) as
   \begin{equation}\label{x1a-x2a}
     x_1 = \frac{k_++k_-}{2},\qquad
     x_2 = \frac{i(k_+ - k_-)}{2},
   \end{equation}
     satisfy the commutation relation
   \begin{equation}\label{commut-x1a,x2a}
      \left[x_1,x_2\right]=\frac{i}{2}(n_1+n_2+1).
   \end{equation}
     The uncertainty relation for these quadratures will be written as follows
   \begin{eqnarray}\label{uncer-x1a-x2a}
      \langle(\Delta x_1)^2\rangle\langle(\Delta x_2)^2\rangle &\geq&
      \frac 1 4 \left|\langle\left[x_1,x_2\right]\rangle\right|^2.
   \end{eqnarray}
     A state is said to exhibit $su(1, 1)$-squeezing if
   \begin{eqnarray}\label{cond-squ-x1a,x2a}
    S_{x_i} = \langle(\Delta x_i)^2\rangle -
     \frac 1 2 |\langle\left[x_1,x_2\right]\rangle| < \;0,\qquad i=1\;  \mbox{or}\;  2.
   \end{eqnarray}

     Following the Liu proposal \cite{liu-quesene} we now consider two hermitian $f-$deformed quadratures as
   \begin{equation}\label{X1-X2}
     X_1 = \frac{K_++K_-}{2},\qquad   X_2 = \frac{i(K_+-K_-)}{2},
   \end{equation}
   where  $K_+$ and $K_-$ were defined in (\ref{nld-generator}).
   These quadratures satisfy the commutation relation
   \begin{equation}\label{commut-X1,X2}
      \left[X_1,X_2\right]=\frac{i}{2}[(n_1+1){f^2(n_1+1)}(n_2+1){f^2(n_2+1)} - n_1{f^2(n_1)}n_2{f^2(n_2)}].
   \end{equation}
    The uncertainty relation for these quadratures will be written as follows
   \begin{eqnarray}\label{uncer-X1-X2}
    \langle(\Delta X_1)^2\rangle\langle(\Delta X_2)^2\rangle &\geq&
    \frac 1 4 \left|\langle\left[X_1,X_2\right]\rangle\right|^2.
   \end{eqnarray}
   A state may exhibit generalized $su(1, 1)$-squeezing if
  \begin{equation}\label{X1-X2}
     S_{X_i} = \langle(\Delta X_i)^2\rangle -
     \frac 1 2 |\langle \left[X_1, X_2\right]\rangle| < \;0,\qquad i=1\;  \mbox{or}\;  2.
   \end{equation}


   \subsection{Quantum statistics of nonlinear charge coherent states}\label{sub-5-1-5}
     A familiar quantity in quantum statistics is the photon count probability.
     The probability of being $n$ photons in first mode and $n-q$ photons in the second mode
     for nonlinear charge coherent states in (\ref{bouth-non-ch-css}) may be given in the
      following closed form expression

    \begin{equation}\label{probability-disturb}
     P\left(n+\frac{q+|q|}{2}, n-\frac{q-|q|}{2}\right)= {\mathcal{N}(|\xi|^2)^{-1}}
      \frac{|\xi|^{2n}}{n!\left[n+|q|\right]!}
     \frac{1} {\{\left[f(n)\right]!\left[f(n+|q|)\right]!\}^2}.
   \end{equation}
     Moreover, there are two measures that may be used to study the fluctuations of quanta number distribution as
     the {"Mandel's $Q$ parameter"} and "second order correlation function" which will be considered by us.

 \subsubsection{Mandel parameter and it's generalization}  \label{sub4-1-5}
    Mandel parameter ordinarily is defined as follows \cite{mandel}:
   \begin{equation}\label{mandel-a}
     Q^a_i=\frac{\langle n_{i}^{2}\rangle-{\langle n_i\rangle}^2}{{\langle n_i\rangle}}-1,
   \end{equation}
     where the   superscript "$a$" indicates that we have used bosonic ladder operators
      and hence $n_i=a_{i}^{\dag} a_i$,
     and the subscript  $i=1,\; 2$ indicates the first and second mode, respectively.

      At this point we would like to  generalize the definition (\ref{mandel-a}) to the following form
   \begin{equation}\label{mandel-with A}
     Q^A_i=\frac{\langle N_{i}^{2}\rangle-{\langle N_i\rangle}^2}{{\langle N_i\rangle}}-1,
   \end{equation}
     where again the  superscript "$A$" indicates that we have used $N_i=A_{i}^{\dag} A$, and the subscript $i=1,\; 2$
     indicates the first and second mode, respectively.

      The negativity of Mandel parameter in (\ref{mandel-a}) or its generalized form in (\ref{mandel-with A})
       indicates that the quantum statistics is  sub-Poissonian, which shows the nonclassicality of the state.

   \subsubsection{Two-mode correlation function and it's
                  generalization} \label{sub4-1-5}
  Generalizing  correlation function definition  of a single-mode field,  Liu defined
   the two-mode correlation function as \cite{liu10}:
    \begin{equation}\label{g factor}
      g=\frac{\langle(n_1n_2)^2\rangle}{{\langle n_1n_2\rangle}^2}.
    \end{equation}
     Two-photon number distribution is related to this measure, where it determines the two-photon
     correlations degree in a two-mode field. If $g<1$ the state is said to have two-mode antibunching
     characterizes the nonclassically of the  state.

     Following Liu  definition in \cite{liu10}  we now  define the "{\it generalized
     two-mode correlation function}" as:
   \begin{equation}\label{gener-g factor}
     G=\frac{\langle(N_1N_2)^2\rangle}{{\langle N_1N_2\rangle}^2},
   \end{equation}
     where $N_i={A_i}^\dag A_i$ for $i=1, 2$ .
     If $G<1$ the state is said to have generalized two-mode antibunching, i.e., it
      possesses  nonclassicality behavior.

    \subsection{Nonclassicality of even and odd nonlinear charge coherent states }
    \label{sec-5-2-5}
      We followed this section with a few words on the nonclassicality of the
      even and odd nonlinear charge coherent states.
      It can be straightforwardly checked that the single- and two-mode squeezing
      (and their generalized forms) may not be occurred for even and odd nonlinear charge
      coherent states. This observations are the same as the previously mentioned results
      for the original nonlinear charge coherent states.
      But, other criteria such as $su(1, 1)$-squeezing, Mandel parameter and the generalized
       form of it introduced respectively in (\ref{mandel-a}) and
      (\ref{mandel-with A}), also the two-mode correlation function as well as it's
      generalized form respectively introduced in (\ref{g factor}) and (\ref{gener-g factor}), will be discussed for
      even (and odd)  nonlinear charge coherent state.

      The probability distributions for even and odd nonlinear charge coherent states are respectively given by
   \begin{equation}\label{ph-disturib-e}
      P_e(2n,2n+|q|)={\mathcal{N}_e(|\xi|^2)}^{-1}\frac{{|\xi|}^{2n}}
      {(2n)!\left[2n+|q|\right]!\{\left[f(2n)\right]!\left[f(2n-q)\right]!\}^2},
   \end{equation}
      and
   \begin{eqnarray}\label{ph-distrib-o}
      P_o(2n+1,2n+1+|q|)&=&{\mathcal{N}_o(|\xi|^2)}^{-1} \{(2n+1)!\left[2n+1+|q|\right]!\}^{-1}\nonumber\\
       &\times &\frac{{|\xi|}^{2n+1}}{\{\left[f(2n+1)\right]!\left[f(2n+1+|q|)\right]!\}^2}.
   \end{eqnarray}
      It is clear that $P_e(2n+1,2n+1+|q|)=0$ and $P_o(2n,2n+|q|)=0$.
      Oscillatory nature of these distributions indicates the obvious  nonclassically  feature of the latter states.

 \section{New classes of nonlinear charge coherent states corresponding to physical systems and their nonclassical features}\label{no-7}

       In order  to illustrate the physical applications of the presented formalism in the paper, let us apply
       the structure on some physical systems which the associated "{\it nonlinear  coherent
       states}" and so their corresponding
       nonlinearity functions, or their "{\it discrete
       spectrum}"  have already been known. Recalling the normal-ordered Hamiltonian of the nonlinear oscillator
       it will be enough for our intention to introduce the discrete spectrum $e_n$ or the nonlinearity function  $f(n)$.
       According to the proposed method by one of us in \cite{Roknizadeh2004, tavassoly-j.m.p} these two physical
       quantities are simply related by $e_n=nf^2(n)$. Thus, knowing the explicit form of $e_n$ one can obtain the
       corresponding nonlinearity function as $f(n)=\sqrt{\frac{e_n}{n}}$ and vice versa. Then, substituting
       a special nonlinearity function  $f(n)$ into "equations (\ref{positive-non-ch-css})
       and (\ref{negative-non-ch-css})" or
       "equations (\ref{e-non-gq}) and (\ref{o-non-gq})" give readily the explicit form of
       "nonlinear charge coherent states"
       or "even and odd nonlinear charge coherent states" associated to each particular system, respectively. Therefore, to economize
       in space the explicit form of nonlinear charge coherent states associated to particular physical systems will
       be introduced in the continuation of the paper have not been given.

       It must be noticed that since we are interested in the nonclassical properties of
       the states of the field, and due
       to the exitance of numerous nonclassical criteria have been outlined in the paper,
       we will mainly focus on the presentation of  the numerical
       results which the nonclassicality aspects of the introduced states may be deeply clarified.

      \subsection{Application to physical systems with known discrete spectra}\label{1-7}

      In this subsection we will deal with some classes of nonlinear coherent states
      associated to quantum systems  whose corresponding
      spectra are known.
      Recall that it contains only a few quantum mechanical systems in physical context.
      Among them we will pay attention to P\"{o}schl-Teller potential and Hydrogen-like spectrum.

       {\it Example 1, P\"{o}schl-Teller potential}:
       The interest in this potential and its coherent states is due to various applications in many fields of physics
       particularly in atomic and molecular physics. The associated Gazeau-Klauder coherent states have been
       demonstrated and discussed nicely by Antonie {\it et al} \cite{antoni} with the convergence  radius   $R=\infty$.
       The corresponding spectrum is $e_n=n(n+\nu)$, so the associated nonlinearity function is \cite{Roknizadeh2004}
      \begin{equation}\label{f-posh-teller}
       f_{PT}(n)=\sqrt{n+\nu},\qquad \nu\geq 2,
      \end{equation}
      where the case $\nu=2$ corresponds to infinite potential well.  We fixed the parameters
      $q=2$ and $\nu=3$ in all computational calculations which
      take into account this particular system. Our numerical results are as follows.
      From figure 1 it is seen that $su(1,1)-$squeezing  in $x_1$ quadrature occurs for
      nonlinear charge coherent states (\ref{positive-non-ch-css}) corresponding to the latter potential.
      While for the standard charge states no such squeezing is
      seen in neither of the quadratures.
      Our further numerical calculations show that the  $su(1, 1)-$squeezing
      for the corresponding nonlinear charge coherent state occurs in $X_1$ quadrature, too.
      Figure 2 shows that for both classes of even charge states, either standard or nonlinear, $su(1, 1)-$squeezing has been
      occurred in $x_2$ quadrature in some regions of $x=|\xi|^2$.
      Figure 3 shows that the same results may be obtained qualitatively for $X_1,\; X_2$.
      But in this case the strength of squeezing in $X_2$ for the nonlinear charge states is much higher than the standard one.
      According to our numerical results concerning the correlation $g-$factor for
      latter potential and all of the introduced nonlinear charge states
      is less than 1. For instance, see figure 4 which indicates the two-mode antibunchig effect for
      the nonlinear charge coherent states as well as the same quantity for the standard charge states.
      It is clear from the figure that while $g <1$ nearly in all
      range of $x$, for the standard charge states it is
      approximated to 1, i.e. the same as canonical (classical)
      coherent states.
      Figure 5  shows the generalized correlation $G-$factor
      for odd nonlinear charge coherent state together with the standard charge states,
      from which we observe that $G<1$ for nonlinear charge states in a wide region of $x$, indicates the nonclassicality
      of the nonlinear odd charge states. This quantity for the standard charge states also becomes negative, but only in a finite
      region of space, i.e., $x<1.75$.
      Mandel parameter in (\ref{mandel-a}) and its generalized form in (\ref{mandel-with A})
      for the nonlinear charge states in both modes can be negative. For instance, as it is shown in figure 6 Mandel
      parameter has been plotted using (\ref{mandel-a}) for first mode, which is negative for arbitrary $x$,
      either standard or nonlinear charge states. But it is
      noticeable that in contrast to all previous cases in which
      the nonclassicality of nonlinear charge states become more high
      light with respect to the standard one, concerning with the latter
      quantity although both are negative, the negativity of the
      nonlinear charge states are less than standard one.


       {\it Example 2, Hydrogen-like spectrum}:
       We now choose the hydrogen-like spectrum whose the corresponding coherent states have been
       a long standing subject and discussed frequently in the literature \cite{Gazeau1}.
       The one-dimensional model of such
       a system with the Hamiltonian $\hat{H}=-\frac{\omega}{({n}+1)^2}$ and the eigenvalues $E_n=-\omega/(n+1)^2$
       has been considered $(\omega=me^4/2,\; \mbox{and}\; n=0, 1, 2, ...)$.
       But to be consistent with the proposed formalism in \cite{tavassoly-j.m.p}
       one must take the shifted Hamiltonian  $e_n=1-\frac 1 {(n+1)^2}$ where we have taken $\omega\equiv1$.
       Thus, for the nonlinearity function corresponding  to such a system one obtains \cite{tavassoly-j.m.p}
      \begin{equation}\label{f-hidrogen}
        f_H(n)=\frac{\sqrt{n+2}}{n+1}.
      \end{equation}
       The nonlinear coherent states corresponding to this nonlinearity function may be defined on the unit
       disk centered at the origin.
       We fixed the charge parameter  $q=2$ in all cases which
       take into account the hydrogen spectrum.
       Our numerical calculations show that the $su(1,1)-$squeezing occurs in $x_2$ quadrature when
       the nonlinear charge coherent states and even nonlinear charge coherent states take into
       account (see for example figure 7 which is plotted
       for nonlinear charge coherent states). It is also seen from
       figure 7 that for the standard charge states the graphs of squeezing coincide with the horizontal axis.
       So nonlinearity causes the nonclassicality considering this
       criteria.
       According to our computational results, for nonlinear charge coherent
       states associated to hydrogen-like atoms the  Heisenberg uncertainty relation
       (\ref{uncer-X1-X2}) is saturated. Therefore, the
       obtained states can be regarded as the {\it generalized intelligent states}
        \cite{parsaiean} with respect to those quadratures.
        This situation also holds
         for standard charge states as it is seen from figure 1.
       For even nonlinear charge coherent states $su(1, 1)-$squeezing will be observed
       in $X_1$ quadratures,  and the same behavior occurs for the standard charge states (see figure
       8).
       For the associated odd nonlinear charge coherent states the inequality $g<1$ holds
       in a finite region of $x$ near  the origin  (see figure 9).
       As it is observed this quantity is approximated to zero for the standard charge states (so possess no nonclassicality criteria).
       Our numerical calculations shows that $G<1$ for all of the permitted values of $x$
       associated to odd nonlinear charge coherent states. Based on our numerical results Mandel parameter $Q_1^a$
       for the first mode of the odd nonlinear charge coherent states is negative in some regions of $x$.
       The parameter $Q^a_2$ for second mode with respect to all of the corresponding classes of charge
       coherent states will be  negative in some regions of $x$.
       The generalized Mandel parameter $Q^A_1$
       will be negative for the first mode of even nonlinear charge coherent states
       while this quantity for the standard charge states is
       approximated to 1 (no classical feature) (see figure 10),
       and for odd one we obtained $Q^A_1 \simeq -1$.
        For second mode we obtained  $Q^A_2\simeq -1$ for all
       introduced classes of corresponding charge states. Recall that Mandel parameter of number states
       as the most nonclassical states is also $-1$.

       To this end, the oscillatory nature of photon distributions for even and odd nonlinear charge coherent
       states associated to both above physical systems is an intrinsic feature of these states,
        so need not to be discussed.

       \subsection{Application to  physical systems with known nonlinearity functions}\label{sub2-7}

       In this subsection we will concern with some classes of nonlinear coherent states whose corresponding
       nonlinearity functions previously introduced in the literature. Particularly they
       are harmonious states and $SU(1, 1)$ coherent states.

        {\it Example 1, Harmonious states}:
         Sudarshan introduced coherent states for simple un-weighted  shift operators that has named harmonious
         states with the nonlinearity function \cite{sudarshan-har.}
        \begin{equation}\label{f-harmonious}
           f_{HS}(n)=\frac 1 {\sqrt{n}}.
        \end{equation}
         Our numerical outputs for nonlinear charge coherent states associated to  this system
         are very closely the same as the results obtained for  hydrogen atom.
         So, paying attention to its dual family \cite{ali, royroy} with
         \begin{equation}\label{f-dual-harmonious}
           f_{DHS}(n)=\sqrt{n},
         \end{equation}
          is preferred.
          This nonlinearity function appears in a natural way in Hamiltonians describing
          interaction with intensity-dependent
          coupling between a two-level atom and an electromagnetic field \cite{singh}.
          We again  observe  that the results
          of our calculations are very closely the same as those of  P\"{o}shl-Teller potential. The latter two facts
          may be expected from comparing the nonlinearity function (\ref{f-harmonious}) with (\ref{f-hidrogen}) and
           (\ref{f-dual-harmonious}) with (\ref{f-posh-teller}).

         {\it Example 2,  $SU(1, 1)$ states}:
         Two distinct sets of generalized coherent states have been introduced in the
         literature associated to $su(1, 1)$ Lie algebra
         known as Barut-Girardello \cite{barut-girardello} and  Gilmore-Perelomov
         \cite{perelomove, gilmore} coherent states.
         In \cite{Roknizadeh2004} it has been demonstrated that Barut-Girardello coherent states can be created via
         the nonlinear coherent states  method by means of nonlinearity function as
        \begin{equation}\label{f-su(1, 1)}
          f_{BG}(n)=\sqrt{n+2\kappa-1},\qquad \kappa=1/2, 1, 3/2, ...\; .
        \end{equation}
         The convergence radius of corresponding states is $R=\infty$. Our numerical results for the corresponding
         nonlinear charge coherent states and their even and odd counterparts  indicate $su(1,1)-$squeezing  with
         respect to nonlinear charge coherent states and even counterpart in $x_2$ quadrature. On the other side,
         for odd nonlinear charge coherent states $su(1, 1)-$squeezing is observed in $X_2$ quadrature.
         For  all of the corresponding charge states $g-$factor is less than 1 showing the nonclassicality of states, for instance see figure 11 which
         has been plotted using the corresponding  nonlinear charge coherent states.  But as it is observed this quantity is also nearly 1 for the standard charge states.
         According to our numerical results $G-$factor
         for associated  nonlinear charge coherent states is $\simeq 1$, while for even and odd ones it takes values
         less than 1.
         For second mode of $SU(1, 1)$ nonlinear charge coherent states, generalized Mandel parameter
         for all of the charge  states and modes will be negative at least in some finite regions (see figure 12 as an example).
         In this last case $Q_A$ becomes negative for both the standard and nonlinear charge states.

        Gilmore-Perelomov coherent states are  defined by the nonlinearity function
       \begin{equation}\label{f-dual-su(1,1)}
         f_{GP}(n)=1/\sqrt{n+2\kappa-1},\quad \kappa=1/2,1, 3/2, ...\;.
       \end{equation}
         The convergence radius of corresponding states is the unit disk. The two sets of states Barut-Girardello
         and Gilmore-Perelomov are known as the dual pair \cite{Roknizadeh2004}.   Our numerical calculations
         show that the corresponding graphical  results are closely the same as the charge states of  hydrogen-like
         spectrum and harmonious states, qualitatively. This may be expected due to the functional form of the
         nonlinearity functions.  The oscillatory nature of photon distribution for all of the associated even
         and odd nonlinear charge coherent states is again a clear fact.

     \section{Summary and conclusion}\label{summary}

       We sum up our presented results as follows.
       Based on the well-known nonlinear coherent states approach in quantum optics filed, we  proposed a formalism
       to introduce the nonlinear charge coherent states in a general framework  associated to any classes of
       "nonlinear coherent states" as well as arbitrary "solvable quantum system" with known  discrete non-degenerate spectra.
       The nonlinear deformed algebra of the nonlinear charge coherent states is realized as $su_Q(1, 1)$.
       After deducing the explicit form of the proposed states, we constructed the even and odd nonlinear
       charge coherent states, through symmetric and antisymmetric superposition of the states, respectively.
       Various  nonclassical properties of all classes of the introduced  states associated to a few quantum
       physical systems have been  investigated, numerically. The dependence of the nonclassicality nature
       on the "nonlinearity function" or "energy spectrum" has been clearly shown.
       According to our numerical results, generally, but not always the nonclassicality will be stronger
       when one deals with nonlinear charge coherent states
       comparing  with the standard charge coherent states.
       It is noticeable that this conclusion depends on the choice of $f(n)$
       and also the specific nonclassicality criteria.
       Besides the nonclassical properties which is the main interest in this field of research, a specific
       aspect of some of the introduced states are noticeable. We may refer to the nonlinear charge coherent state
       corresponding to hydrogen-like spectrum which behaves as intelligent state
       in $X_1$ and $X_2$ quadratures.
       To this end, our formalism provides a general and simple scheme for the construction of a vast classes
       of charge coherent states and their even and odd counterparts associated to any particular nonlinearity
       function as well as arbitrary solvable quantum system with discrete spectra.
       For instance, the application of the formalism
       to photon-added and depleted coherent states \cite{sivakumar}, anharmonic oscillator coherent states
       \cite{popov}, isotonic oscillator \cite{honarasa, mkthonarasa2}
       and other nonlinear oscillator systems, and investigating their nonclassical properties is a straightforward
       matter may be done elsewhere. We hope that the  nonlinear charge coherent states and their superpositions
       have been introduced in the present paper will also find their applications in various physical fields,
       as well as the standard charge coherent states \cite{paricle1}-\cite{Botke}.

  \newpage
    \vspace {2 cm}
   {\bf FIGURE CAPTIONS}
    \vspace {.5 cm}

    {\bf FIG. 1.} The plot of $su(1, 1)$-squeezing for $x_1$ quadrature and  $x_2$ quadrature
    as a function of $x=|\xi|^2$ for standard charge coherent state ($f(n)=1$), and
    nonlinear charge coherent state associated to
    P\"{o}shl-Teller potential ($f_{PT}(n)$) with fixed parameters
    $\nu = 3, q=2$. For the case $f(n)=1$ the  continues   curve with point   and
    dot-dashed curve, and for the case $f_{PT}(n)$ the dashed curve and solid curve are
    respectively indicate  $x_1$ and $x_2$ $su(1, 1)$-squeezing.
    Notice that the continues   curve with points and dot-dashed curve coincide with each other in the horizontal axis.

   {\bf FIG. 2.} The same as Fig. 1 for even nonlinear charge
    coherent state associated to standard charge coherent states ($f(n)=1$) and
    P\"{o}shl-Teller potential ($f_{PT}(n)$) with fixed parameters
    $\nu = 3, q=2$. For the case $f(n)=1$ the  dot-dashed curve and
    continues   curve with points, and for the case $f_{PT}(n)$ the dashed curve and solid curve are
    respectively indicate  $x_1$ and $x_2$ $su(1, 1)$-squeezing.

   {\bf FIG. 3.} The same as Fig. 2 except that the $X_1$ and
   $X_2$ quadratures are considered.

    {\bf FIG. 4.} The plot of $g$-factor as a function of
     $x=|\xi|^2$ for standard charge coherent state ($f(n)=1$), and
    nonlinear charge coherent state associated to
    P\"{o}shl-Teller potential ($f_{PT}(n)$) with fixed parameters
    $\nu = 3, q=2$. The dashed curve corresponds to the case $f(n)=1$ and the
    solid curve corresponds to the case $f_{PT}(n)$.

    {\bf FIG. 5.} The plot of $G$-factor as a function of
     $x=|\xi|^2$ for standard charge coherent state ($f(n)=1$), and
    nonlinear charge coherent state associated to
    P\"{o}shl-Teller potential ($f_{PT}(n)$) with fixed parameters
    $\nu = 3, q=2$. The dashed curve corresponds to the case $f(n)=1$ and the
    solid curve corresponds to the case $f_{PT}(n)$.

    {\bf FIG. 6.} The graph of Mandel parameter for the first
    mode as a function of $x=|\xi|^2$ for standard charge coherent state ($f(n)=1$), and
    nonlinear charge coherent state associated to
    P\"{o}shl-Teller potential ($f_{PT}(n)$) with fixed parameters
    $\nu = 3, q=2$. The dashed curve corresponds to the case $f(n)=1$ and the
    solid curve corresponds to the case $f_{PT}(n)$.

   {\bf FIG. 7.} The same as Fig. 2 except that
    standard charge coherent state ($f(n)=1$), and
    nonlinear charge coherent state associated to
    hydrogen-like spectrum  ($f_{H}(n)$) is considered with fixed parameter $q=2$.

   {\bf FIG. 8.} The same as Fig. 3  except that
    standard charge coherent state ($f(n)=1$), and
    nonlinear charge coherent state associated to
    hydrogen-like spectrum  ($f_{H}(n)$) is considered with fixed parameter $q=2$.

    {\bf FIG. 9.} The same as Fig. 4  except that
    standard odd charge coherent state ($f(n)=1$), and
    odd nonlinear charge coherent state associated to
    hydrogen-like spectrum  ($f_{H}(n)$) is considered with fixed parameter $q=2$.

    {\bf FIG. 10.} The same as Fig. 6 the graph of generalized Mandel parameter for the second
    mode as a function of $x=|\xi|^2$ for
    standard even  charge coherent state ($f(n)=1$), and
    even nonlinear charge coherent state associated to
    hydrogen-like spectrum  ($f_{H}(n)$) is considered with fixed parameter $q=2$.

   {\bf FIG. 11.} The plot of $g$-factor as a function of
     $x=|\xi|^2$ for standard charge coherent state ($f(n)=1$), and
    nonlinear charge coherent state associated to
    Barut-Girardello  charge coherent states  ($f_{BG}(n)$) with fixed parameters
    $\chi=1/2, q=2$. The dashed curve corresponds to the case $f(n)=1$ and the
    solid curve corresponds to the case $f_{PT}(n)$.

   {\bf FIG. 12.} The graph of generalized Mandel parameter for the second
    mode as a function of $x=|\xi|^2$ for standard charge coherent state ($f(n)=1$), and
    nonlinear charge coherent state associated to
    Barut-Girardello ($f_{BG}(n)$) with fixed parameters
    $\chi=1/2, q=2$. The dashed curve corresponds to the case $f(n)=1$ and the
    solid curve corresponds to the case $f_{BG}(n)$.

 \end{document}